\documentclass[aps,prc,twocolumn,superscriptaddress,showpacs,floatfix]{revtex4-1}
\usepackage{graphicx,amssymb}
\usepackage[fleqn]{amsmath}
\usepackage{amsfonts}
\usepackage{dcolumn}
\usepackage{bm}
\usepackage{braket}
\usepackage{multirow} 
\usepackage{MnSymbol}

\graphicspath{{images/}}

\newcommand{\hatvec}[1]
{\hat{\vec{#1}}}

\renewcommand{\vec}[1]{\mbox{\boldmath $#1$}}

\renewcommand{\vec}[1]{\mbox{\boldmath $#1$}}

\begin{document}

\date{\today}

\title{Continuum effects in neutron-drip-line oxygen isotopes}

\author{K. Fossez}
\affiliation{NSCL/FRIB Laboratory,
Michigan State University, East Lansing, Michigan 48824, USA}

\author{J. Rotureau}
\affiliation{NSCL/FRIB Laboratory,
Michigan State University, East Lansing, Michigan 48824, USA}
\affiliation{JINPA, Oak Ridge National Laboratory, Oak Ridge, TN 37831, USA}

\author{N. Michel}
\affiliation{NSCL/FRIB Laboratory,
Michigan State University, East Lansing, Michigan 48824, USA}

\author{W. Nazarewicz}
\affiliation{Department of Physics and Astronomy and FRIB Laboratory,
Michigan State University, East Lansing, Michigan 48824, USA}

\begin{abstract}
The binding-energy pattern along the neutron-rich oxygen chain, governed by an interplay between shell effects and many-body correlations impacted by strong couplings to one- and two-neutron continuum, make these isotopes a unique testing ground for nuclear models. In this work, we investigate ground states and low-lying excited states of $^{23-28}$O using the complex-energy Gamow Shell Model and Density Matrix Renormalization Group method with a finite-range two-body interaction optimized to the bound states and resonances of $^{23-26}$O, assuming a core of $^{22}$O. Our results suggest that the ground-state of $^{28}$O has a threshold character, i.e., is very weakly bound or slightly unbound. 
We also predict narrow excited resonances in $^{25}$O and $^{27}$O. 
The inclusion of the large continuum space significantly impacts predicted binding energies of $^{26-28}$O. This implies that the careful treatment of neutron continuum is necessary prior to assessing the spectroscopic quality of effective interactions in this region.
\end{abstract}

\maketitle

{\it Introduction} --- 
The neutron-rich oxygen isotopes ${ {}^{23-28}\text{O} }$ constitute an excellent laboratory for the study of an interplay between single-particle motion and many-body correlations in the presence of neutron continuum \cite{forssen13_394}.
The semi-magic character of oxygen isotopes makes the shell model picture fairly robust up to ${ {}^{24}\text{O} }$, 
with $^{22}$O corresponding to the $\nu (0d_{5/2})^6$ subshell closure \cite{thirolf00_1783,stanoiu04_1774,elekes06_1785,becheva06_1784}, 
bound	$^{23}$O\cite{elekes07_1768,schiller07_1769}, 
and $^{24}$O associated with the $\nu (0d_{5/2})^6(1s_{1/2})^2$ subshell closure \cite{hoffman09_1350,hoffman11_1331,tshoo12_1771}. 

According to the current experimental evidence, the neutron drip line for $Z=8$ is reached at ${ {}^{24}\text{O} }$, which is believed to be the last bound oxygen isotope. Indeed, the isotope $^{25}$O has been shown to be unbound \cite{hoffman08_1781,caesar13_1765} 
as well as the two-neutron emitter $^{26}$O \cite{guillemaud90_1761,lunderberg12_556} 
which appears to be a very narrow threshold resonance \cite{kohley13_1541,caesar13_1765,kondo16_1439}.
While the odd-$N$ isotope $^{27}$O is believed to be unbound \cite{sakurai99_1754}, the situation is far from clear for $^{28}$O
as in a shell model (SM) picture its apparent doubly-magic character could in principle result in an enhanced stability. 
Experimentally, there have been hints \cite{doornenbal17_1850} of
the reduced neutron magicity toward $^{28}$O, 
and several measurements \cite{tarasov97_1786,sakurai99_1754} have provided circumstantial evidence for the unbound character of this nucleus. 
However, in the absence of direct measurement, the jury is still out on the question of how much unbound this system really is.

\begin{figure}[htb]
	\includegraphics[width=1.0\linewidth]{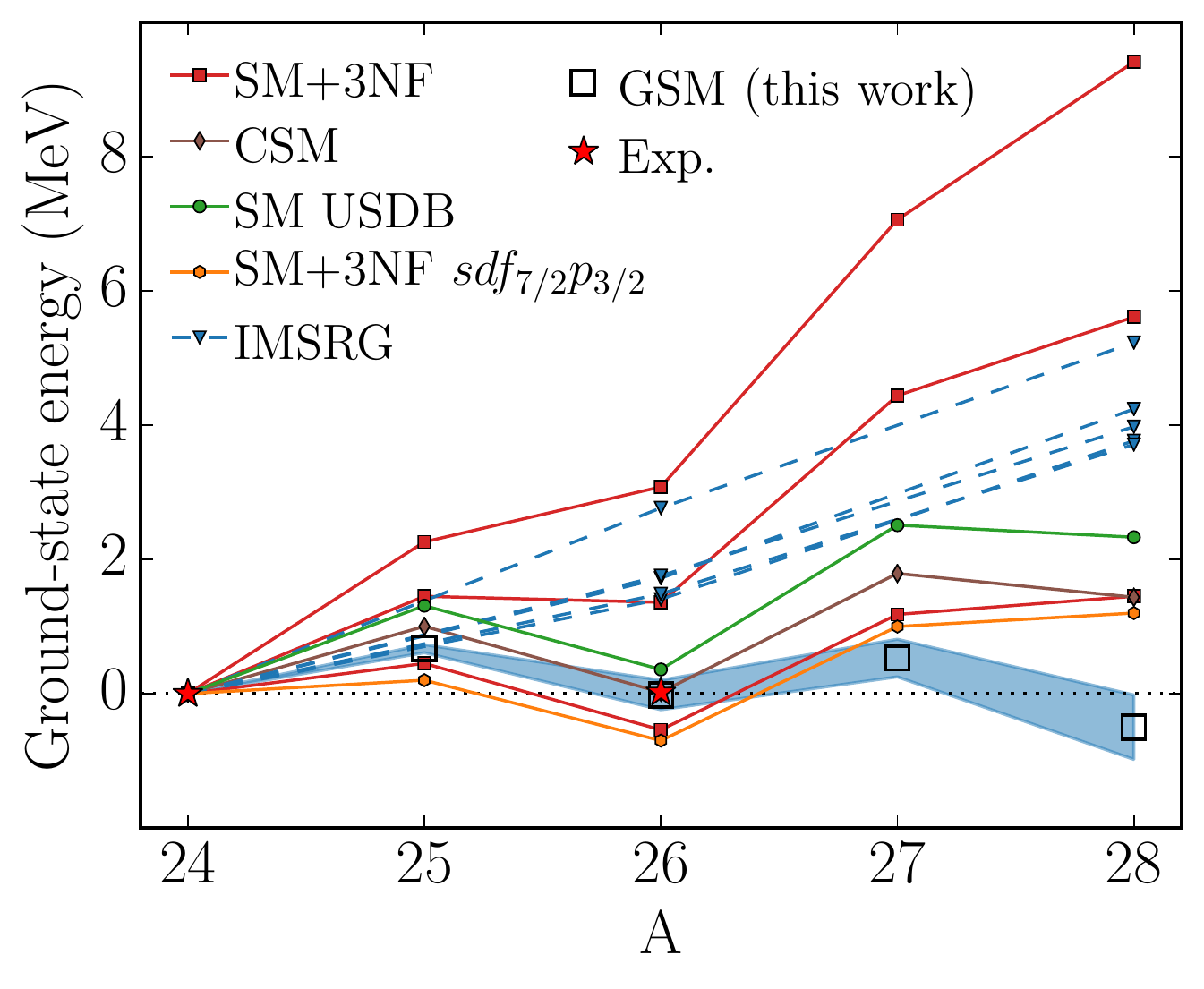}
	\caption{Ground-state binding energies of $^{25-28}$O relative to $^{24}$O obtained in various theoretical approaches using various interactions:
	SM USDB \cite{brown17_1876}, SM+3NF \cite{otsuka10_1027}, SM+3NF in the extended ${ sd{f}_{7/2}{p}_{3/2} }$ space, CSM \cite{volya05_470,volya06_94}, and
	IMSRG \cite{lapoux16_1782,privcom}. If a given theoretical approach was used with several interactions, those predictions are marked by identical symbols/lines. 
	Open boxes represent the results obtained in this work (variant B) and the shaded area shows the impact of the enlarged continuum space, see discussion in the text.}
	\label{fig1}
\end{figure}

On the theory side, 
various many-body approaches using realistic chiral nucleon-nucleon (NN) interactions investigated the stability and structure of $^{23-26}$O by considering the neutron continuum space \cite{tsukiyama09_497,hagen12_685,tsukiyama15_1442,hagen16_1800,sun17_1840}. However, since the interactions used were not fine-tuned to experiment
and the continuum model spaces were severely truncated, only qualitative predictions 
were made for energies and widths of unbound states.
Concerning $^{28}$O, 
early SM calculations \cite{caurier98_1790,utsuno01_1789,brown17_1876} 
predicted $^{28}$O to be two-neutron unstable due to the unbound character of the ${ 0{d}_{3/2} }$ single particle (s.p.) shell.
The inclusion of the continuum space and related couplings within the continuum SM (CSM) \cite{kruppa97_485,volya05_470,volya06_94} also yielded $^{28}$O outside the two-neutron drip line.
Early many-body investigations using realistic chiral interactions produced more nuanced results, predicting $^{28}$O bound or unbound, depending on the renormalization cutoff of the interaction in the absence of the continuum space \cite{hagen09_1778}, and computing it largely unbound when continuum was included \cite{hagen12_685}. 
The role of 3NF in neutron-rich oxygen isotopes was first investigated using SM approaches and renormalized chiral interactions \cite{otsuka10_1027,caesar13_1765,holt13_1871}, and it was concluded that repulsive 3NF could lead to the decreased stability of the
heaviest oxygen isotopes.
Finally, state-of-the-art calculations including chiral two- and three-body forces provided consistent results along the neutron-rich oxygen chain 
\cite{hergert13_1718,hergert14_1719,bogner14_1156,cipollone13_1857,jansen14_1150,epelbaum14_1893,binder14_1882,cipollone15_1894,hergert17_1873,stroberg17_1807,arxivTichai17}, 
demonstrating that at a given level of approximation all methods are under control. 
However, in the absence of continuum couplings, these predictions deviate from experiment for the heaviest isotopes known, e.g., $^{26}$O.
(An overview of the recent progresses can be found in Refs.~\cite{hebeler15_1872,stroberg17_1807}.)
The current situation in the neutron-rich oxygen isotopes is summarized in Fig.~\ref{fig1}, 
which shows g.s. binding energies of $^{25-28}$O relative to $^{24}$O predicted in various models.
The chiral-interaction results are represented by the IM-SRG predictions \cite{lapoux16_1782,privcom} obtained using various forces. 
The appreciable spread between various theoretical predictions for the g.s. energy of $^{26-28}$O provides a strong motivation for a consistent microscopic description of neutron-rich isotopes using a realistic model optimized locally to data on oxygen isotopes, and fully including the couplings due to the neutron continuum. The latter is critical as the weakly bound/unbound oxygen isotopes are prototypical open quantum systems \cite{michel10_4}.

{\it Theoretical framework} ---
Earlier work on the neutron-rich oxygen isotopes demonstrated that their behavior results from the subtle balance between many-body dynamics, realistic forces, and continuum coupling.
In the present study, following the strategy of Ref.~\cite{tsukiyama09_497}, we choose to investigate $^{25-28}$O in the configuration-interaction picture by considering a core of $^{22}$O and by optimizing a realistic NN interaction to the experimentally known states in ${ {}^{23-26}\text{O} }$. We use the Gamow Shell Model (GSM) \cite{michel09_2}, which is an extension of the traditional SM into the complex-energy plane through the use of the Berggren ensemble \cite{berggren68_32,berggren82_28}. The advantage of GSM is that it can describe many-body bound states and resonances
within one consistent framework. 
By adjusting the parameters of the GSM Hamiltonian to neutron-rich isotopes we hope to absorb effectively the leading 3NF effects while making a ``minimal" extrapolation in neutron number from $N=18$ to $N=20$.

The key element of GSM is the s.p. Berggren ensemble \cite{berggren68_32}, which explicitly includes bound states, decaying resonances, and non-resonant scattering continuum \cite{michel09_2}. The GSM approach formally allows to describe an arbitrary number of valence nucleons in the continuum, but
there are practical limitations if large configuration spaces are involved, as 
the particle continuum needs to be discretized. 
Some of those limitations can be tamed by limiting the maximal number of particles that can occupy continuum shells for a given configuration. Such truncations can often be justified when many-body correlations in the continuum are not playing a major role. However, when configurations involving several nucleons in the continuum are essential, as, e.g., in two-neutron emitters such as the g.s. of $^{26}$O, another approach is needed. One possible way to avoid the explosion of the configuration space due to the discretized continuum 
is the use of the Density Matrix Renormalization Group (DMRG) method \cite{rotureau06_15,rotureau09_140} where the continuum couplings are included progressively. The general idea of the DMRG approach is to start from a truncated many-body space, which provides a first approximation to the eigenstate in the full space, and then to gradually add scattering states while retaining those many-body states that provide the largest contribution to the GSM density matrix.

The GSM Hamiltonian used in this work contains the kinetic energy ${ \hat{t} }$ of valence nucleons, one- and two-body potentials ${ \hat{U} }$ and ${ \hat{V} }$, respectively, as well as a recoil term that guarantees translational invariance. In order to estimate the impact of the core on predictions, we extended the GSM by including configurations containing a single neutron hole in $0{d}_{5/2}$. In order to prevent double counting, we removed from the GSM Hamiltonian 
the one-body potential ${ \hat{W}_{i} }$ that represents the interaction between the ${ i }$-th valence neutron and
 the $\nu{ 0{d}_{5/2} }$ shell.
The renormalized GSM Hamiltonian that can account for a one-neutron hole configuration can thus be written as:
\begin{equation}
	\hat{H} - \hat{H}_{ \text{core} } = \sum_{i = 1}^{ {N}_{ \text{val} } + 1 } ( \hat{t}_{i} + \hat{U}_{i} - \hat{W}_{i} ) + \sum_{ i < k }^{ {N}_{ \text{val} } + 1 } \left( \hat{V}_{ik} + \frac{ \hatvec{p}_{i}.\hatvec{p}_{k} }{ {M}_{ \text{core} } } \right),
	\label{eq_final_H}
\end{equation}
where ${ {N}_{ \text{val} } }$ is the number of valence particles and ${ {M}_{ \text{core} } }$ is the core mass. This treatment of the hole configuration is referred to as `the GSM-hole' in the following. In deriving (\ref{eq_final_H}) we assume that the core potential is not altered when a hole is present in the hole shell.

{\it Optimized interaction} --- 
The main objective of this study is to provide reliable predictions on neutron-rich oxygen isotopes that will guide experiment and benchmark future calculations by estimating the impact of continuum couplings on heaviest oxygen isotopes. To optimize the interaction, we maximize the number of experimental data points constraining the GSM Hamiltonian while minimizing the number of valence particles considered in order to deal with a reasonable model space. A good compromise is obtained for a core of ${ {}^{22}\text{O} }$. The corresponding s.p. model space is limited to the bound ${ 1{s}_{1/2} }$ and resonant ${ 0{d}_{3/2} }$ shells and associated scattering continua, each made of three segments in the complex momentum plane defined by the points ${ (0.15,0.0) }$, ${ (0.3,0.0) }$ and ${ (2.0,0.0) }$ (in ${ \text{fm}^{-1} }$) for the ${ {s}_{1/2} }$ partial wave, and ${ (0.25,-0.05) }$, ${ (0.5,0.0) }$ and ${ (2.0,0.0) }$ (in ${ \text{fm}^{-1} }$) for the ${ {d}_{3/2} }$ partial wave.
We take 15 points for the $s_{1/2}$ contour and 24 points for the $d_{3/2}$ contour. During optimization, we retain configurations with at most three neutrons in the scattering shells; this slightly affects the fit for $^{26}$O but allows tractable calculations. 
By fitting the two-body interaction as well as the core potential one effectively absorbs the most important part of 3NF.

As shown in Table~\ref{tab_fit}, there are nine experimentally known states in ${ {}^{23-26}\text{O} }$. The ${ {J}^{ \pi } = {5/2}^{+} }$ state in $^{23}$O is interpreted as a ${ 0{d}_{5/2} }$ hole configuration; it can only be described in a GSM-hole picture. The energy of this state provides a useful constraint on the core potential. 
\begin{table}[htb]
	\caption{Experimental energies (in MeV) and widths (in keV) of $^{23-26}$O \cite{ensdf} compared to the results of interactions A and B employed in this work. }
	\begin{ruledtabular}
		\begin{tabular}{cclll}
			Nucleus & ${ {J}^{\pi} }$ & ${ E_{\text{exp}} }$ & $E_{\rm A}$ & $E_{\rm B}$ \\
			\hline \\[-6pt]
			\multirow{3}{*}
			{$^{23}$O} 	& ${ {1/2}^{+} }$ & $-2.74$			& $-2.67$ 	& $-2.59$ 	\\
		 			& ${ {5/2}^{+} }$ & 0.06 ($\sim$0) 		& 0.04 		& 		\\
					& ${ {3/2}^{+} }$ & 1.26 ($\lesssim$1.3)	& 1.29 (227)	& 1.32 (241) 	\\
			\hline \\[-6pt]
			\multirow{3}{*}
			{$^{24}$O} 	& ${ {0}^{+} }$ & $-6.35$ 			& $-6.34$	& $-6.29$ 	\\
					& ${ {2}^{+} }$ & $-1.64 (\lesssim 0.63)$ 	& $-1.97 (56)$ 	& $-1.87 (60)$ 	\\
					& ${ {1}^{+} }$ & $-1.03 (\lesssim 1.24)$ 	& $-1.19 (311)$ & $-1.09 (312)$ \\
			\hline \\[-6pt]
			{$^{25}$O} 	& ${ {3/2}^{+} }$ & $-5.60 (\lesssim 0.77)$ 	& $-5.68 (48)$ 	& $-5.62 (51)$ \\
			\hline \\[-6pt]
			\multirow{2}{*}
			{$^{26}$O} 	& ${ {0}^{+} }$ & $-6.33 (\lesssim 0.1)$ 	& $-6.39$ 	& $-6.31$ 	\\
					& ${ {2}^{+} }$ & $-5.05 (\lesssim 0.5)$ 	& $-5.32 (26)$ 	& $-5.23 (27)$ 	
		\end{tabular}
	\end{ruledtabular}
	\label{tab_fit}
\end{table}

The widths were not included in the fit for two reasons.
First, experimental widths are not well known; in general they are overestimated. Second, energies and widths of Gamow states are correlated, which makes the optimization algorithm unstable.
In general, low-lying unbound states in neutron-rich oxygen isotopes are expected to have small decay widths that are primarily dominated by the ${ {d}_{3/2} }$ partial wave.
Our predicted widths are in fact consistent with other calculations \cite{volya06_94,grigorenko13_1219,grigorenko15_1874,hove17_1893}.

The energy of the excited $J^\pi = 1^+$ state in $^{24}$O has not been considered in the optimization since it does not constrain the fit when the $J^\pi = 2^+$ state energy is included. 
The optimization of the core potential can be better achieved within the GSM-hole technique since it allows to include the important ${ {J}^{ \pi } = {5/2}^{+} }$ level in $^{23}$O in the fit. The core-neutron interaction is represented by a Woods-Saxon potential defined as in Ref.~\cite{michel09_2}, with the fixed diffuseness $a$=0.65\,fm and radius ${R}_{0}$=3.15\,fm. The adjustable parameters are: the $\ell$-dependent strengths ${ {V}_{0}^{(\ell)} }$ and spin-orbit strength ${ {V}_{ \text{so} } }$. The effective interaction is the Furutani-Horiuchi-Tamagaki (FHT) finite-range two-body interaction \cite{furutani78_1012,furutani79_1013}, which is described by means of 4 free parameters when only neutrons are considered. Altogether our GSM Hamiltonian contains 7 parameters that are constrained by 8 known states in $^{23-26}$O. Once the initial optimization with GSM-hole is done (variant A), the core potential parameters are frozen and employed in the second optimization (variant B) without a 
$\nu 0{d}_{5/2}$ hole involved. Variant B involves 4 FHT interaction parameters constrained to 7 experimental levels.

\begin{figure}[htb]
	\includegraphics[width=1.0\linewidth]{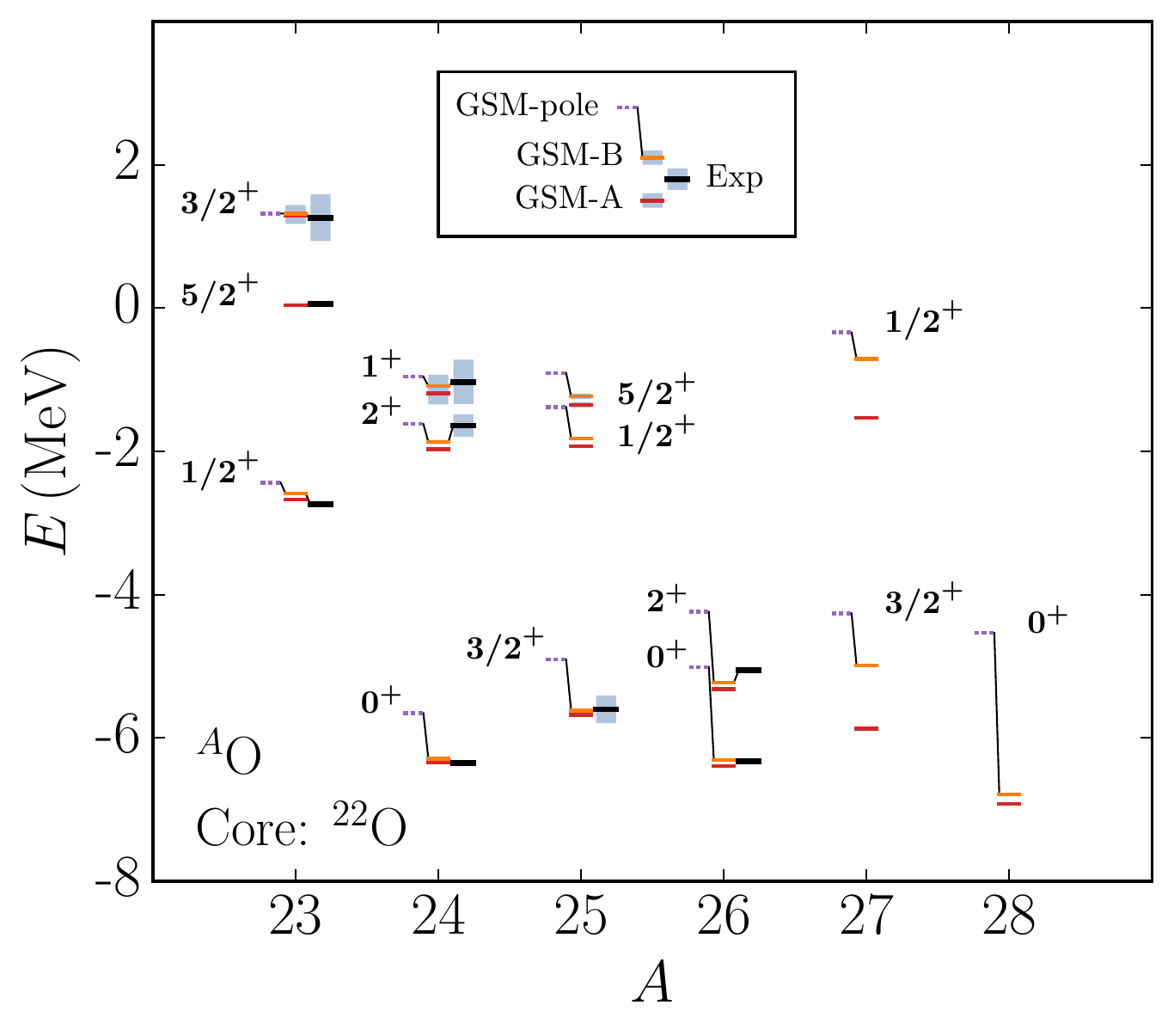}
	\caption{Energies of $^{23-28}$O obtained in GSM variants A and B and compared to experiment. The ``GSM-pole" results are obtained by using the interaction of variant B and removing the scattering space. The widths are marked by shaded bands.}
	\label{fig2}
\end{figure}

The parameters of the Woods-Saxon potential obtained in variant A are: ${ {V}_{0}^{(\ell = 0)} = 50.1 \, \text{MeV} }$, ${ {V}_{0}^{(\ell = 2)} = 54.1 \, \text{MeV} }$, 
and $V_{ \text{so} } = 9.27$\,MeV. The FTH interaction parameters are listed in Table~\ref{tab_inter}.
\begin{table}[htb]
	\caption{Strengths of the central (c), spin-orbit (so), and tensor (t) terms of the FHT interaction in different spin-isospin $(S,T)$ channels obtained in various optimization variants. The values of
		${ {V}_{ \text{c} }^{S,T} }$ and ${ {V}_{ \text{so} }^{S,T} }$ are in MeV and ${ {V}_{ \text{t} }^{S,T} }$ is in ${ \text{MeV} \, \text{fm}^{-2} }$.}
	\begin{ruledtabular}
		\begin{tabular}{lcccc}
			Variant & ${ {V}_{\text{c}}^{1,1} }$ & ${ {V}_{\text{c}}^{0,1} }$ & ${ {V}_{\text{so}}^{1,1} }$ & ${ {V}_{\text{t}}^{1,1} }$ \\
			\hline\\[-7pt]
			A & -$9.49$ & $-4.16$ & $-240.1$ & 15.2 \\
			B & $-9.21$ & $-4.13$ & $-237.1$ & 14.8
		\end{tabular}
	\end{ruledtabular}
	\label{tab_inter}
\end{table}
The energies of $^{23-26}$O obtained in different optimization variants are displayed in Table~\ref{tab_fit} and shown in Fig.~\ref{fig2}. Because the GSM-hole approximation involves a one-body potential which depends on the two-body interaction, the levels in $^{23}$O are predicted to be slightly different in variants A and B. It is seen that the predicted energy of the $J^\pi = 1^+$ state in $^{24}$O is close to experiment in all cases. The optimizations A and B are quite well constrained and result in similar interaction parameters; hence, both yield comparable reproduction of experimental data. This suggests that -- when it comes to predictions for $^{27,28}$O -- there is no advantage in using a model space including configurations containing a $\nu 0{d}_{5/2}$ hole.

To see whether our predictions can be affected by the configuration-space truncation, we used the DMRG method, which enables calculations with no restriction on the number of valence neutrons in scattering shells.
Removing the scattering-state truncations also provides a better estimate of particle widths because the many-body completeness relation is met.
\begin{table}[htb]
	\caption{Predictions for energies (in MeV) and widths (in keV) obtained in GSM and DMRG using interaction B. The convergence criterion in DMRG \cite{rotureau06_15,rotureau09_140} is $\varepsilon = 10^{-8} $.}
	\begin{ruledtabular}
		\begin{tabular}{ccll}
			Nucleus & ${ {J}^{\pi} }$ & ${ {E}_{ \text{GSM} } }$ & ${ {E}_{ \text{DMRG} } }$\\
			\hline \\[-6pt]

			\multirow{3}{*}
			{$^{25}$O}	& ${ {3/2}^{+} }$ 	& $-5.62 (51)$	\\
					& ${ {1/2}^{+} }$ & 			 $-1.82 (0) $	\\
					& ${ {5/2}^{+} }$ & 			 $-1.23 (79)$	\\
			\hline \\[-6pt]
			\multirow{2}{*}
			{$^{26}$O} 	& ${ {0}^{+} }$ 	& $-$6.31 (0) 	& $-$6.30 (0) 		\\
					& ${ {2}^{+} }$	& $-$5.23 (27) & $-$5.22 (10) 		\\
			\hline \\[-6pt]
			\multirow{2}{*}
			{$^{27}$O} 	& ${ {3/2}^{+} }$ &		 $-$5.76 (14) & $-$5.76 ($<$10) 	\\
					& ${ {1/2}^{+} }$ & 		 $-$1.42 (17) & $-$1.43 ($<$10) 	\\
			\hline \\[-6pt]
			$^{28}$O 	& ${ {0}^{+} }$ & 	 $-$6.79 	& $-$6.74			
		\end{tabular}
	\end{ruledtabular}
	\label{tab_predict}
\end{table}
As seen in Table~\ref{tab_predict}, the DMRG results are practically identical to the original GSM results, which demonstrates the validity of the assumed GSM truncations.

{\it Results} --- 
Our predictions for energies of $^{25,27,28}$O and without restriction on the number of neutrons in the continuum are shown in Fig.~\ref{fig2}. 
It is seen that the g.s. of $^{28}$O is calculated to be bound by about 500 keV in both variants A and B, which is in apparent contradiction with the current experimental evidence \cite{tarasov97_1786,sakurai99_1754}. 
In order to show the effect of the scattering continuum, we show the GSM results obtained with the interaction B by removing the space of scatering states (``GSM pole"). While such results are obtained in an incomplete space, they still provide a fair assessment of the continuum coupling. When compared to the GSM-B calculation, we conclude that the g.s. of $^{24}$O and $^{28}$O gain about 500 keV and 2 MeV of binding energy, respectively, by including the continuum space. We note that when removing continuum couplings the g.s. energies of $^{27}$O and $^{28}$O show a similar trend as those in the SM results. 

A factor that can impact our predictions is the uncertainty of interaction parameters.
In fact, our optimization study indicates that
the FHT parameters ${ {V}_{ \text{c} }^{1,1} }$ and ${ {V}_{\text{so}}^{1,1} }$ are poorly constrained by the adopted dataset.
However, by varying these ``sloppy" coupling constants around the optimization minimum it was impossible to reproduce the experimental g.s. of $^{26}$O and unbound $^{28}$O.

\begin{figure}[htb]
	\includegraphics[width=1.0\linewidth]{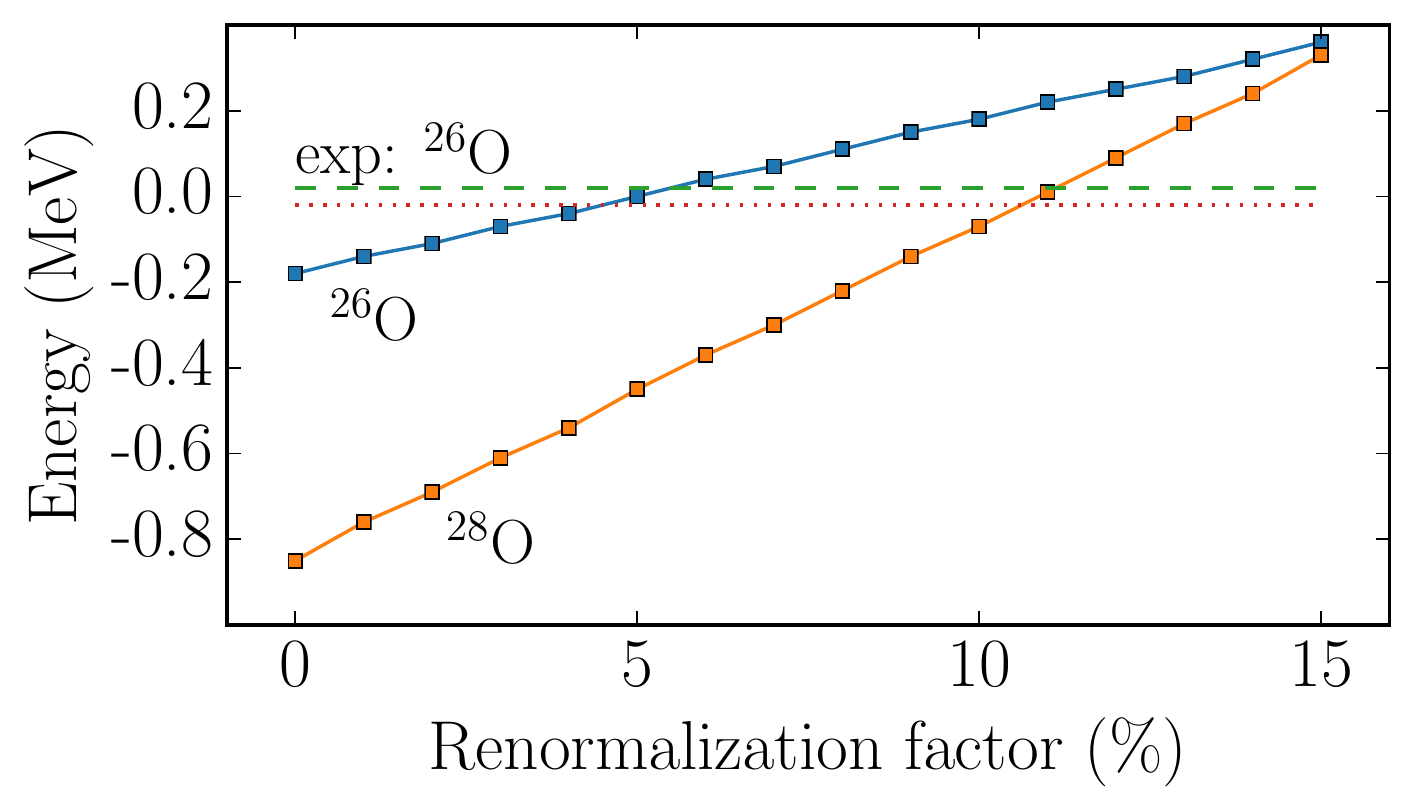}
	\caption{Ground-state energies of $^{26}$O and $^{28}$O relative to $^{24}$O computed in the extended $spdf$ space as a function of the renormalization factor for the interaction B. The $sd$-space prediction for $^{26}$O is marked by a dotted line.}
	\label{fig3}
\end{figure}

We also investigated the impact of the increased model space. Based on the earlier 
studies of dineutron correlations, we know that 
threshold systems such as $^{26}$O are strongly affected by couplings between positive and negative parity states \cite{catara84_1880,pillet07_1879,hagino14_1160,hagino16_1881}.
Consequently, extending model space by including $p$ and $f$ waves could improve our predictions for $^{26-28}$O. Using the interaction of variant B as a starting point, we computed $^{23-28}$O in DMRG in a larger model space 
including the $p_{3/2}$ and $p_{1/2}$ scattering continua defined by the points ${ (0.25,-0.05) }$, ${ (0.5,0.0) }$ and ${ (2.0,0.0) }$ (in ${ \text{fm}^{-1} }$) in the complex momentum plane 
and each discretized by 24 scattering states. We also included
the $d_{5/2}$, $f_{7/2}$ and $f_{5/2}$ real-energy continua, each represented by 5 harmonic oscillator shells. 
In order to compensate for the growth of the model space, we decreased the strength of the FHT interaction by introducing an overall renormalization factor ranging between 0 and 15\%.
Figure~\ref{fig3}
shows the resulting evolution of the g.s. energies of $^{26}$O and $^{28}$O with respect to the g.s. energy of $^{24}$O. The latter value turned out to be very weakly affected by the increased model space. This is not surprising as the bound g.s. of $^{24}$O is not expected to be strongly affected by dineutron correlations.

The enlarged model space results in an increased binding of $^{26}$O (by about 200\,keV) and $^{28}$O (by about 350\,keV) due to the continuum coupling. By decreasing the interaction strength by $\sim$5\%, one can bring the g.s. energy of $^{26}$O back to the experiment value; this results in $S_{\rm 2n}\approx 400$\,keV for $^{28}$O.
The two-neutron threshold in $^{28}$O decreases with the renormalization factor: the 4n threshold is reached at a $\sim$11\% renormalization and $^{26,28}$O are predicted to become 2n-unbound by about 300\,keV
at a $\sim$15\% renormalization. We note that a similar outcome for $^{28}$O is obtained in the $sd$-space GSM calculations based on the reoptimized FHT interaction using a dataset with the experimental g.s. energy of $^{26}$O shifted by 300\,keV.

An estimate of the uncertainties due to the $pf$-continuum couplings missing in $sd$-space calculations 
is obtained by the energy change of the g.s. in $^{25-28}$O when renormalizing the interaction until the g.s. of $^{26}$O reaches the experimental value.
Such a procedure results in shifts of 0.07 MeV, 0.25 MeV, 0.32 MeV, and 0.55 MeV, for the g.s. energies of $^{25-28}$O, respectively; these uncertainties are represented in Fig.~\ref{fig1} by a shaded area.
While our analysis predicts the g.s. of $^{28}$O to be more likely bound than unbound, it also indicates that this system has a threshold character, with 2n and 4n thresholds being close in energy.

An interesting prediction of this work is the possible existence of narrow states in $^{25}$O and $^{27}$O with dominant SM configurations based on the $1s_{1/2}$ neutron hole. 
For those states, the neutron emission is governed by the ${ \ell = 2 }$ waves, which results in small decay widths of about 80 keV for the ${ {J}^{\pi} = {5/2}^{+} }$ state of $^{25}$O and less than 10 keV for the other states.
The energies of excited states in $^{25}$O are robust with respect to changes in the interaction, while those in $^{27}$O exhibit appreciable variations.
Still, for all the interaction variants considered, there is a possibility for the ${ {J}^{\pi} = {1/2}^{+} }$ state of $^{27}$O to decay through the emission of one- and two neutrons. 
No excited states in $^{25}$O were observed in single-proton removal reaction from $^{26}$F \cite{kondo16_1439}. This is not surprising, as the $1s_{1/2}$ neutron hole character of those states would result in a small spectroscopic factor. The decay pattern of the ${ {J}^{\pi} = {3/2}^{+} }$ g.s. of $^{27}$O is sensitive to the interaction used. According to variant A, this state is expected exhibit a neutron decay to the g.s. of $^{26}$O. In variant B, a one-neutron branch to the 2$^+$ state of $^{26}$O and a two-neutron branch to the g.s. of $^{25}$O are also predicted.

{\it Conclusions} ---
The structure of neutron-rich oxygen isotopes was investigated within the GSM+DMRG framework, which 
enables the description of many-body dynamics in the presence of continuum couplings.
Assuming a core of $^{22}$O, we optimized the finite-range FHT interaction to the bound states and resonances of $^{23-26}$O. In this way, our predictions for $^{27,28}$O are based on a minimal extrapolation in neutron number.
According to our model, the g.s. of $^{28}$O has a threshold character, i.e., it is either weakly bound or weakly unbound within the uncertainty of our approach. Another prediction concerns
the possible existence of excited narrow neutron ${ {J}^{\pi} = {1/2}^{+} }$ resonances in $^{25,27}$O, 
with dominant SM configurations involving a $1s_{1/2}$ neutron hole. 

The impact of the non-resonant continuum on theoretical predictions for $^{26-28}$O is appreciable. 
In particular, 
the effect of the $sp$ continuum space on the g.s. energy of $^{28}$O is $\sim$2\,MeV; by adding $pf$ scattering continua, g.s. energy is further decreased by 0.55\,MeV. Clearly, without considering large continuum spaces and guaranteeing consistency with measured energies, it is difficult to make a definitive statement on the missing pieces of the interaction, e.g., the role of repulsive effective 3NF. 
In summary, we believe that our GSM+DMRG predictions provide a strong motivation for further experimental and theoretical investigations in this region.


\begin{acknowledgments}
Useful discussions with Michael 
Thoennessen and Alexandra Gade are gratefully acknowledged.
This work was supported by the U.S.\ Department of Energy, Office of
Science, Office of Nuclear Physics under award numbers 
DE-SC0013365 (Michigan State University) and DE-SC0008511 (NUCLEI SciDAC-3 collaboration), and by the National Science Foundation under award number PHY-1403906.
\end{acknowledgments}


\begin{thebibliography}{63}%
\makeatletter
\providecommand \@ifxundefined [1]{%
 \@ifx{#1\undefined}
}%
\providecommand \@ifnum [1]{%
 \ifnum #1\expandafter \@firstoftwo
 \else \expandafter \@secondoftwo
 \fi
}%
\providecommand \@ifx [1]{%
 \ifx #1\expandafter \@firstoftwo
 \else \expandafter \@secondoftwo
 \fi
}%
\providecommand \natexlab [1]{#1}%
\providecommand \enquote  [1]{``#1''}%
\providecommand \bibnamefont  [1]{#1}%
\providecommand \bibfnamefont [1]{#1}%
\providecommand \citenamefont [1]{#1}%
\providecommand \href@noop [0]{\@secondoftwo}%
\providecommand \href [0]{\begingroup \@sanitize@url \@href}%
\providecommand \@href[1]{\@@startlink{#1}\@@href}%
\providecommand \@@href[1]{\endgroup#1\@@endlink}%
\providecommand \@sanitize@url [0]{\catcode `\\12\catcode `\$12\catcode
  `\&12\catcode `\#12\catcode `\^12\catcode `\_12\catcode `\%12\relax}%
\providecommand \@@startlink[1]{}%
\providecommand \@@endlink[0]{}%
\providecommand \url  [0]{\begingroup\@sanitize@url \@url }%
\providecommand \@url [1]{\endgroup\@href {#1}{\urlprefix }}%
\providecommand \urlprefix  [0]{URL }%
\providecommand \Eprint [0]{\href }%
\providecommand \doibase [0]{http://dx.doi.org/}%
\providecommand \selectlanguage [0]{\@gobble}%
\providecommand \bibinfo  [0]{\@secondoftwo}%
\providecommand \bibfield  [0]{\@secondoftwo}%
\providecommand \translation [1]{[#1]}%
\providecommand \BibitemOpen [0]{}%
\providecommand \bibitemStop [0]{}%
\providecommand \bibitemNoStop [0]{.\EOS\space}%
\providecommand \EOS [0]{\spacefactor3000\relax}%
\providecommand \BibitemShut  [1]{\csname bibitem#1\endcsname}%
\let\auto@bib@innerbib\@empty
\bibitem [{\citenamefont {Forss\'en}\ \emph {et~al.}(2013)\citenamefont
  {Forss\'en}, \citenamefont {Hagen}, \citenamefont {{Hjorth-Jensen}},
  \citenamefont {Nazarewicz},\ and\ \citenamefont {Rotureau}}]{forssen13_394}%
  \BibitemOpen
  \bibfield  {author} {\bibinfo {author} {\bibfnamefont {C.}~\bibnamefont
  {Forss\'en}}, \bibinfo {author} {\bibfnamefont {G.}~\bibnamefont {Hagen}},
  \bibinfo {author} {\bibfnamefont {M.}~\bibnamefont {{Hjorth-Jensen}}},
  \bibinfo {author} {\bibfnamefont {W.}~\bibnamefont {Nazarewicz}}, \ and\
  \bibinfo {author} {\bibfnamefont {J.}~\bibnamefont {Rotureau}},\ }\href
  {https://dx.doi.org/10.1088/0031-8949/2013/T152/014022} {\bibfield  {journal}
  {\bibinfo  {journal} {Phys. Scr. T}\ }\textbf {\bibinfo {volume} {152}},\
  \bibinfo {pages} {014022} (\bibinfo {year} {2013})}\BibitemShut {NoStop}%
\bibitem [{\citenamefont {Thirolf}\ \emph {et~al.}(2000)\citenamefont {Thirolf}
  \emph {et~al.}}]{thirolf00_1783}%
  \BibitemOpen
  \bibfield  {author} {\bibinfo {author} {\bibfnamefont {P.~G.}\ \bibnamefont
  {Thirolf}} \emph {et~al.},\ }\href
  {http://dx.doi.org/10.1016/S0370-2693(00)00720-6} {\bibfield  {journal}
  {\bibinfo  {journal} {Phys. Lett. B}\ }\textbf {\bibinfo {volume} {485}},\
  \bibinfo {pages} {16} (\bibinfo {year} {2000})}\BibitemShut {NoStop}%
\bibitem [{\citenamefont {Stanoiu}\ \emph {et~al.}(2004)\citenamefont {Stanoiu}
  \emph {et~al.}}]{stanoiu04_1774}%
  \BibitemOpen
  \bibfield  {author} {\bibinfo {author} {\bibfnamefont {M.}~\bibnamefont
  {Stanoiu}} \emph {et~al.},\ }\href
  {http://dx.doi.org/10.1103/PhysRevC.69.034312} {\bibfield  {journal}
  {\bibinfo  {journal} {Phys. Rev. C}\ }\textbf {\bibinfo {volume} {69}},\
  \bibinfo {pages} {034312} (\bibinfo {year} {2004})}\BibitemShut {NoStop}%
\bibitem [{\citenamefont {Elekes}\ \emph {et~al.}(2006)\citenamefont {Elekes}
  \emph {et~al.}}]{elekes06_1785}%
  \BibitemOpen
  \bibfield  {author} {\bibinfo {author} {\bibfnamefont {Z.}~\bibnamefont
  {Elekes}} \emph {et~al.},\ }\href
  {http://dx.doi.org/10.1103/PhysRevC.74.017306} {\bibfield  {journal}
  {\bibinfo  {journal} {Phys. Rev. C}\ }\textbf {\bibinfo {volume} {74}},\
  \bibinfo {pages} {017306} (\bibinfo {year} {2006})}\BibitemShut {NoStop}%
\bibitem [{\citenamefont {Becheva}\ \emph {et~al.}(2006)\citenamefont {Becheva}
  \emph {et~al.}}]{becheva06_1784}%
  \BibitemOpen
  \bibfield  {author} {\bibinfo {author} {\bibfnamefont {E.}~\bibnamefont
  {Becheva}} \emph {et~al.},\ }\href
  {http://dx.doi.org/10.1103/PhysRevLett.96.012501} {\bibfield  {journal}
  {\bibinfo  {journal} {Phys. Rev. Lett.}\ }\textbf {\bibinfo {volume} {96}},\
  \bibinfo {pages} {012501} (\bibinfo {year} {2006})}\BibitemShut {NoStop}%
\bibitem [{\citenamefont {Elekes}\ \emph {et~al.}(2007)\citenamefont {Elekes}
  \emph {et~al.}}]{elekes07_1768}%
  \BibitemOpen
  \bibfield  {author} {\bibinfo {author} {\bibfnamefont {Z.}~\bibnamefont
  {Elekes}} \emph {et~al.},\ }\href
  {http://dx.doi.org/10.1103/PhysRevLett.98.102502} {\bibfield  {journal}
  {\bibinfo  {journal} {Phys. Rev. Lett.}\ }\textbf {\bibinfo {volume} {98}},\
  \bibinfo {pages} {102502} (\bibinfo {year} {2007})}\BibitemShut {NoStop}%
\bibitem [{\citenamefont {Schiller}\ \emph {et~al.}(2007)\citenamefont
  {Schiller} \emph {et~al.}}]{schiller07_1769}%
  \BibitemOpen
  \bibfield  {author} {\bibinfo {author} {\bibfnamefont {A.}~\bibnamefont
  {Schiller}} \emph {et~al.},\ }\href
  {http://dx.doi.org/10.1103/PhysRevLett.99.112501} {\bibfield  {journal}
  {\bibinfo  {journal} {Phys. Rev. Lett.}\ }\textbf {\bibinfo {volume} {99}},\
  \bibinfo {pages} {112501} (\bibinfo {year} {2007})}\BibitemShut {NoStop}%
\bibitem [{\citenamefont {Hoffman}\ \emph {et~al.}(2009)\citenamefont {Hoffman}
  \emph {et~al.}}]{hoffman09_1350}%
  \BibitemOpen
  \bibfield  {author} {\bibinfo {author} {\bibfnamefont {C.~R.}\ \bibnamefont
  {Hoffman}} \emph {et~al.},\ }\href
  {http://dx.doi.org/10.1016/j.physletb.2008.12.066} {\bibfield  {journal}
  {\bibinfo  {journal} {Phys. Lett. B}\ }\textbf {\bibinfo {volume} {672}},\
  \bibinfo {pages} {17} (\bibinfo {year} {2009})}\BibitemShut {NoStop}%
\bibitem [{\citenamefont {Hoffman}\ \emph {et~al.}(2011)\citenamefont {Hoffman}
  \emph {et~al.}}]{hoffman11_1331}%
  \BibitemOpen
  \bibfield  {author} {\bibinfo {author} {\bibfnamefont {C.~R.}\ \bibnamefont
  {Hoffman}} \emph {et~al.},\ }\href
  {http://dx.doi.org/10.1103/PhysRevC.83.031303} {\bibfield  {journal}
  {\bibinfo  {journal} {Phys. Rev. C}\ }\textbf {\bibinfo {volume} {83}},\
  \bibinfo {pages} {031303(R)} (\bibinfo {year} {2011})}\BibitemShut {NoStop}%
\bibitem [{\citenamefont {Tshoo}\ \emph {et~al.}(2012)\citenamefont {Tshoo}
  \emph {et~al.}}]{tshoo12_1771}%
  \BibitemOpen
  \bibfield  {author} {\bibinfo {author} {\bibfnamefont {K.}~\bibnamefont
  {Tshoo}} \emph {et~al.},\ }\href
  {http://dx.doi.org/10.1103/PhysRevLett.109.022501} {\bibfield  {journal}
  {\bibinfo  {journal} {Phys. Rev. Lett.}\ }\textbf {\bibinfo {volume} {109}},\
  \bibinfo {pages} {022501} (\bibinfo {year} {2012})}\BibitemShut {NoStop}%
\bibitem [{\citenamefont {Hoffman}\ \emph {et~al.}(2008)\citenamefont {Hoffman}
  \emph {et~al.}}]{hoffman08_1781}%
  \BibitemOpen
  \bibfield  {author} {\bibinfo {author} {\bibfnamefont {C.~R.}\ \bibnamefont
  {Hoffman}} \emph {et~al.},\ }\href
  {http://dx.doi.org/10.1103/PhysRevLett.100.152502} {\bibfield  {journal}
  {\bibinfo  {journal} {Phys. Rev. Lett.}\ }\textbf {\bibinfo {volume} {100}},\
  \bibinfo {pages} {152502} (\bibinfo {year} {2008})}\BibitemShut {NoStop}%
\bibitem [{\citenamefont {Caesar}\ \emph {et~al.}(2013)\citenamefont {Caesar}
  \emph {et~al.}}]{caesar13_1765}%
  \BibitemOpen
  \bibfield  {author} {\bibinfo {author} {\bibfnamefont {C.}~\bibnamefont
  {Caesar}} \emph {et~al.},\ }\href
  {http://dx.doi.org/10.1103/PhysRevC.88.034313} {\bibfield  {journal}
  {\bibinfo  {journal} {Phys. Rev. C}\ }\textbf {\bibinfo {volume} {88}},\
  \bibinfo {pages} {034313} (\bibinfo {year} {2013})}\BibitemShut {NoStop}%
\bibitem [{\citenamefont {{Guillemaud-Mueller}}\ \emph
  {et~al.}(1990)\citenamefont {{Guillemaud-Mueller}} \emph
  {et~al.}}]{guillemaud90_1761}%
  \BibitemOpen
  \bibfield  {author} {\bibinfo {author} {\bibfnamefont {D.}~\bibnamefont
  {{Guillemaud-Mueller}}} \emph {et~al.},\ }\href
  {http://dx.doi.org/10.1103/PhysRevC.41.937} {\bibfield  {journal} {\bibinfo
  {journal} {Phys. Rev. C}\ }\textbf {\bibinfo {volume} {41}},\ \bibinfo
  {pages} {937} (\bibinfo {year} {1990})}\BibitemShut {NoStop}%
\bibitem [{\citenamefont {Lunderberg}\ \emph {et~al.}(2012)\citenamefont
  {Lunderberg} \emph {et~al.}}]{lunderberg12_556}%
  \BibitemOpen
  \bibfield  {author} {\bibinfo {author} {\bibfnamefont {E.}~\bibnamefont
  {Lunderberg}} \emph {et~al.},\ }\href
  {https://dx.doi.org/10.1103/PhysRevC.68.014612} {\bibfield  {journal}
  {\bibinfo  {journal} {Phys. Rev. Lett.}\ }\textbf {\bibinfo {volume} {108}},\
  \bibinfo {pages} {142503} (\bibinfo {year} {2012})}\BibitemShut {NoStop}%
\bibitem [{\citenamefont {Kohley}\ \emph {et~al.}(2013)\citenamefont {Kohley}
  \emph {et~al.}}]{kohley13_1541}%
  \BibitemOpen
  \bibfield  {author} {\bibinfo {author} {\bibfnamefont {Z.}~\bibnamefont
  {Kohley}} \emph {et~al.},\ }\href
  {http://dx.doi.org/10.1103/PhysRevLett.110.152501} {\bibfield  {journal}
  {\bibinfo  {journal} {Phys. Rev. Lett.}\ }\textbf {\bibinfo {volume} {110}},\
  \bibinfo {pages} {152501} (\bibinfo {year} {2013})}\BibitemShut {NoStop}%
\bibitem [{\citenamefont {Kondo}\ \emph {et~al.}(2016)\citenamefont {Kondo}
  \emph {et~al.}}]{kondo16_1439}%
  \BibitemOpen
  \bibfield  {author} {\bibinfo {author} {\bibfnamefont {Y.}~\bibnamefont
  {Kondo}} \emph {et~al.},\ }\href
  {http://dx.doi.org/10.1103/PhysRevLett.116.102503} {\bibfield  {journal}
  {\bibinfo  {journal} {Phys. Rev. Lett.}\ }\textbf {\bibinfo {volume} {116}},\
  \bibinfo {pages} {102503} (\bibinfo {year} {2016})}\BibitemShut {NoStop}%
\bibitem [{\citenamefont {Sakurai}\ \emph {et~al.}(1999)\citenamefont {Sakurai}
  \emph {et~al.}}]{sakurai99_1754}%
  \BibitemOpen
  \bibfield  {author} {\bibinfo {author} {\bibfnamefont {H.}~\bibnamefont
  {Sakurai}} \emph {et~al.},\ }\href
  {http://dx.doi.org/10.1016/S0370-2693(99)00015-5} {\bibfield  {journal}
  {\bibinfo  {journal} {Phys. Lett. B}\ }\textbf {\bibinfo {volume} {448}},\
  \bibinfo {pages} {180} (\bibinfo {year} {1999})}\BibitemShut {NoStop}%
\bibitem [{\citenamefont {Doornenbal}\ \emph {et~al.}(2017)\citenamefont
  {Doornenbal} \emph {et~al.}}]{doornenbal17_1850}%
  \BibitemOpen
  \bibfield  {author} {\bibinfo {author} {\bibfnamefont {P.}~\bibnamefont
  {Doornenbal}} \emph {et~al.},\ }\href
  {https://doi.org/10.1103/PhysRevC.95.041301} {\bibfield  {journal} {\bibinfo
  {journal} {Phys. Rev. C}\ }\textbf {\bibinfo {volume} {95}},\ \bibinfo
  {pages} {041301(R)} (\bibinfo {year} {2017})}\BibitemShut {NoStop}%
\bibitem [{\citenamefont {Tarasov}\ \emph {et~al.}(1997)\citenamefont {Tarasov}
  \emph {et~al.}}]{tarasov97_1786}%
  \BibitemOpen
  \bibfield  {author} {\bibinfo {author} {\bibfnamefont {O.}~\bibnamefont
  {Tarasov}} \emph {et~al.},\ }\href
  {http://dx.doi.org/10.1016/S0370-2693(97)00901-5} {\bibfield  {journal}
  {\bibinfo  {journal} {Phys. Lett. B}\ }\textbf {\bibinfo {volume} {409}},\
  \bibinfo {pages} {64} (\bibinfo {year} {1997})}\BibitemShut {NoStop}%
\bibitem [{\citenamefont {Tsukiyama}\ \emph {et~al.}(2009)\citenamefont
  {Tsukiyama}, \citenamefont {{Hjorth-Jensen}},\ and\ \citenamefont
  {Hagen}}]{tsukiyama09_497}%
  \BibitemOpen
  \bibfield  {author} {\bibinfo {author} {\bibfnamefont {K.}~\bibnamefont
  {Tsukiyama}}, \bibinfo {author} {\bibfnamefont {M.}~\bibnamefont
  {{Hjorth-Jensen}}}, \ and\ \bibinfo {author} {\bibfnamefont {G.}~\bibnamefont
  {Hagen}},\ }\href {https://dx.doi.org/10.1103/PhysRevC.80.051301} {\bibfield
  {journal} {\bibinfo  {journal} {Phys. Rev. C}\ }\textbf {\bibinfo {volume}
  {80}},\ \bibinfo {pages} {051301(R)} (\bibinfo {year} {2009})}\BibitemShut
  {NoStop}%
\bibitem [{\citenamefont {Hagen}\ \emph {et~al.}(2012)\citenamefont {Hagen},
  \citenamefont {{Hjorth-Jensen}}, \citenamefont {Jansen}, \citenamefont
  {Machleidt},\ and\ \citenamefont {Papenbrock}}]{hagen12_685}%
  \BibitemOpen
  \bibfield  {author} {\bibinfo {author} {\bibfnamefont {G.}~\bibnamefont
  {Hagen}}, \bibinfo {author} {\bibfnamefont {M.}~\bibnamefont
  {{Hjorth-Jensen}}}, \bibinfo {author} {\bibfnamefont {G.~R.}\ \bibnamefont
  {Jansen}}, \bibinfo {author} {\bibfnamefont {R.}~\bibnamefont {Machleidt}}, \
  and\ \bibinfo {author} {\bibfnamefont {T.}~\bibnamefont {Papenbrock}},\
  }\href {https://dx.doi.org/10.1103/PhysRevLett.108.242501} {\bibfield
  {journal} {\bibinfo  {journal} {Phys. Rev. Lett.}\ }\textbf {\bibinfo
  {volume} {108}},\ \bibinfo {pages} {242501} (\bibinfo {year}
  {2012})}\BibitemShut {NoStop}%
\bibitem [{\citenamefont {Tsukiyama}\ \emph {et~al.}(2015)\citenamefont
  {Tsukiyama}, \citenamefont {Otsuka},\ and\ \citenamefont
  {Fujimoto}}]{tsukiyama15_1442}%
  \BibitemOpen
  \bibfield  {author} {\bibinfo {author} {\bibfnamefont {K.}~\bibnamefont
  {Tsukiyama}}, \bibinfo {author} {\bibfnamefont {T.}~\bibnamefont {Otsuka}}, \
  and\ \bibinfo {author} {\bibfnamefont {R.}~\bibnamefont {Fujimoto}},\ }\href
  {https://dx.doi.org/10.1093/ptep/ptv125} {\bibfield  {journal} {\bibinfo
  {journal} {Prog. Theor. Exp. Phys.}\ }\textbf {\bibinfo {volume} {2015}},\
  \bibinfo {pages} {093D01} (\bibinfo {year} {2015})}\BibitemShut {NoStop}%
\bibitem [{\citenamefont {Hagen}\ \emph {et~al.}(2016)\citenamefont {Hagen},
  \citenamefont {{Hjorth-Jensen}}, \citenamefont {Jansen},\ and\ \citenamefont
  {Papenbrock}}]{hagen16_1800}%
  \BibitemOpen
  \bibfield  {author} {\bibinfo {author} {\bibfnamefont {G.}~\bibnamefont
  {Hagen}}, \bibinfo {author} {\bibfnamefont {M.}~\bibnamefont
  {{Hjorth-Jensen}}}, \bibinfo {author} {\bibfnamefont {G.~R.}\ \bibnamefont
  {Jansen}}, \ and\ \bibinfo {author} {\bibfnamefont {T.}~\bibnamefont
  {Papenbrock}},\ }\href {http://dx.doi.org/10.1088/0031-8949/91/6/063006}
  {\bibfield  {journal} {\bibinfo  {journal} {Phys. Scr.}\ }\textbf {\bibinfo
  {volume} {91}},\ \bibinfo {pages} {063006} (\bibinfo {year}
  {2016})}\BibitemShut {NoStop}%
\bibitem [{\citenamefont {Sun}\ \emph {et~al.}(2017)\citenamefont {Sun},
  \citenamefont {Wu}, \citenamefont {Zhao}, \citenamefont {Hu}, \citenamefont
  {Dai},\ and\ \citenamefont {Xu}}]{sun17_1840}%
  \BibitemOpen
  \bibfield  {author} {\bibinfo {author} {\bibfnamefont {Z.~H.}\ \bibnamefont
  {Sun}}, \bibinfo {author} {\bibfnamefont {Q.}~\bibnamefont {Wu}}, \bibinfo
  {author} {\bibfnamefont {Z.~H.}\ \bibnamefont {Zhao}}, \bibinfo {author}
  {\bibfnamefont {B.~S.}\ \bibnamefont {Hu}}, \bibinfo {author} {\bibfnamefont
  {S.~J.}\ \bibnamefont {Dai}}, \ and\ \bibinfo {author} {\bibfnamefont
  {F.~R.}\ \bibnamefont {Xu}},\ }\href
  {http://dx.doi.org/10.1016/j.physletb.2017.03.054} {\bibfield  {journal}
  {\bibinfo  {journal} {Phys. Lett. B}\ }\textbf {\bibinfo {volume} {769}},\
  \bibinfo {pages} {227} (\bibinfo {year} {2017})}\BibitemShut {NoStop}%
\bibitem [{\citenamefont {Caurier}\ \emph {et~al.}(1998)\citenamefont
  {Caurier}, \citenamefont {Nowacki}, \citenamefont {Poves},\ and\
  \citenamefont {Retamosa}}]{caurier98_1790}%
  \BibitemOpen
  \bibfield  {author} {\bibinfo {author} {\bibfnamefont {E.}~\bibnamefont
  {Caurier}}, \bibinfo {author} {\bibfnamefont {F.}~\bibnamefont {Nowacki}},
  \bibinfo {author} {\bibfnamefont {A.}~\bibnamefont {Poves}}, \ and\ \bibinfo
  {author} {\bibfnamefont {J.}~\bibnamefont {Retamosa}},\ }\href
  {http://dx.doi.org/10.1103/PhysRevC.58.2033} {\bibfield  {journal} {\bibinfo
  {journal} {Phys. Rev. C}\ }\textbf {\bibinfo {volume} {58}},\ \bibinfo
  {pages} {2033} (\bibinfo {year} {1998})}\BibitemShut {NoStop}%
\bibitem [{\citenamefont {Utsuno}\ \emph {et~al.}(2001)\citenamefont {Utsuno},
  \citenamefont {Otsuka}, \citenamefont {Mizusaki},\ and\ \citenamefont
  {Honma}}]{utsuno01_1789}%
  \BibitemOpen
  \bibfield  {author} {\bibinfo {author} {\bibfnamefont {Y.}~\bibnamefont
  {Utsuno}}, \bibinfo {author} {\bibfnamefont {T.}~\bibnamefont {Otsuka}},
  \bibinfo {author} {\bibfnamefont {T.}~\bibnamefont {Mizusaki}}, \ and\
  \bibinfo {author} {\bibfnamefont {M.}~\bibnamefont {Honma}},\ }\href
  {http://dx.doi.org/10.1103/PhysRevC.64.011301} {\bibfield  {journal}
  {\bibinfo  {journal} {Phys. Rev. C}\ }\textbf {\bibinfo {volume} {64}},\
  \bibinfo {pages} {011301(R)} (\bibinfo {year} {2001})}\BibitemShut {NoStop}%
\bibitem [{\citenamefont {Brown}(2017)}]{brown17_1876}%
  \BibitemOpen
  \bibfield  {author} {\bibinfo {author} {\bibfnamefont {B.~A.}\ \bibnamefont
  {Brown}},\ }\href {http://dx.doi.org/10.1142/S0218301317400031} {\bibfield
  {journal} {\bibinfo  {journal} {Int. J. Mod. Phys. E}\ }\textbf {\bibinfo
  {volume} {26}},\ \bibinfo {pages} {1740003} (\bibinfo {year}
  {2017})}\BibitemShut {NoStop}%
\bibitem [{\citenamefont {Kruppa}\ \emph {et~al.}(1997)\citenamefont {Kruppa},
  \citenamefont {Heenen}, \citenamefont {Flocard},\ and\ \citenamefont
  {Liotta}}]{kruppa97_485}%
  \BibitemOpen
  \bibfield  {author} {\bibinfo {author} {\bibfnamefont {A.~T.}\ \bibnamefont
  {Kruppa}}, \bibinfo {author} {\bibfnamefont {P.~H.}\ \bibnamefont {Heenen}},
  \bibinfo {author} {\bibfnamefont {H.}~\bibnamefont {Flocard}}, \ and\
  \bibinfo {author} {\bibfnamefont {R.~J.}\ \bibnamefont {Liotta}},\ }\href
  {https://dx.doi.org/10.1103/PhysRevLett.79.2217} {\bibfield  {journal}
  {\bibinfo  {journal} {Phys. Rev. Lett.}\ }\textbf {\bibinfo {volume} {79}},\
  \bibinfo {pages} {2217} (\bibinfo {year} {1997})}\BibitemShut {NoStop}%
\bibitem [{\citenamefont {Volya}\ and\ \citenamefont
  {Zelevinsky}(2005)}]{volya05_470}%
  \BibitemOpen
  \bibfield  {author} {\bibinfo {author} {\bibfnamefont {A.}~\bibnamefont
  {Volya}}\ and\ \bibinfo {author} {\bibfnamefont {V.}~\bibnamefont
  {Zelevinsky}},\ }\href {https://dx.doi.org/10.1103/PhysRevLett.94.052501}
  {\bibfield  {journal} {\bibinfo  {journal} {Phys. Rev. Lett.}\ }\textbf
  {\bibinfo {volume} {94}},\ \bibinfo {pages} {052501} (\bibinfo {year}
  {2005})}\BibitemShut {NoStop}%
\bibitem [{\citenamefont {Volya}\ and\ \citenamefont
  {Zelevinsky}(2006)}]{volya06_94}%
  \BibitemOpen
  \bibfield  {author} {\bibinfo {author} {\bibfnamefont {A.}~\bibnamefont
  {Volya}}\ and\ \bibinfo {author} {\bibfnamefont {V.}~\bibnamefont
  {Zelevinsky}},\ }\href {https://dx.doi.org/10.1103/PhysRevC.74.064314}
  {\bibfield  {journal} {\bibinfo  {journal} {Phys. Rev. C}\ }\textbf {\bibinfo
  {volume} {74}},\ \bibinfo {pages} {064314} (\bibinfo {year}
  {2006})}\BibitemShut {NoStop}%
\bibitem [{\citenamefont {Hagen}\ \emph {et~al.}(2009)\citenamefont {Hagen},
  \citenamefont {Papenbrock}, \citenamefont {Dean}, \citenamefont
  {{Hjorth-Jensen}},\ and\ \citenamefont {{Velamur Asokan}}}]{hagen09_1778}%
  \BibitemOpen
  \bibfield  {author} {\bibinfo {author} {\bibfnamefont {G.}~\bibnamefont
  {Hagen}}, \bibinfo {author} {\bibfnamefont {T.}~\bibnamefont {Papenbrock}},
  \bibinfo {author} {\bibfnamefont {D.~J.}\ \bibnamefont {Dean}}, \bibinfo
  {author} {\bibfnamefont {M.}~\bibnamefont {{Hjorth-Jensen}}}, \ and\ \bibinfo
  {author} {\bibfnamefont {B.}~\bibnamefont {{Velamur Asokan}}},\ }\href
  {http://dx.doi.org/10.1103/PhysRevC.80.021306} {\bibfield  {journal}
  {\bibinfo  {journal} {Phys. Rev. C}\ }\textbf {\bibinfo {volume} {80}},\
  \bibinfo {pages} {021306(R)} (\bibinfo {year} {2009})}\BibitemShut {NoStop}%
\bibitem [{\citenamefont {Otsuka}\ \emph {et~al.}(2010)\citenamefont {Otsuka},
  \citenamefont {Suzuki}, \citenamefont {Holt}, \citenamefont {Schwenk},\ and\
  \citenamefont {Akaishi}}]{otsuka10_1027}%
  \BibitemOpen
  \bibfield  {author} {\bibinfo {author} {\bibfnamefont {T.}~\bibnamefont
  {Otsuka}}, \bibinfo {author} {\bibfnamefont {T.}~\bibnamefont {Suzuki}},
  \bibinfo {author} {\bibfnamefont {J.~D.}\ \bibnamefont {Holt}}, \bibinfo
  {author} {\bibfnamefont {A.}~\bibnamefont {Schwenk}}, \ and\ \bibinfo
  {author} {\bibfnamefont {Y.}~\bibnamefont {Akaishi}},\ }\href
  {https://dx.doi.org/10.1103/PhysRevLett.105.032501} {\bibfield  {journal}
  {\bibinfo  {journal} {Phys. Rev. Lett.}\ }\textbf {\bibinfo {volume} {105}},\
  \bibinfo {pages} {032501} (\bibinfo {year} {2010})}\BibitemShut {NoStop}%
\bibitem [{\citenamefont {Holt}\ \emph {et~al.}(2013)\citenamefont {Holt},
  \citenamefont {Men\'endez},\ and\ \citenamefont {Schwenk}}]{holt13_1871}%
  \BibitemOpen
  \bibfield  {author} {\bibinfo {author} {\bibfnamefont {J.~D.}\ \bibnamefont
  {Holt}}, \bibinfo {author} {\bibfnamefont {J.}~\bibnamefont {Men\'endez}}, \
  and\ \bibinfo {author} {\bibfnamefont {A.}~\bibnamefont {Schwenk}},\ }\href
  {https://doi.org/10.1140/epja/i2013-13039-2} {\bibfield  {journal} {\bibinfo
  {journal} {Eur. Phys. J. A}\ }\textbf {\bibinfo {volume} {49}},\ \bibinfo
  {pages} {39} (\bibinfo {year} {2013})}\BibitemShut {NoStop}%
\bibitem [{\citenamefont {Hergert}\ \emph {et~al.}(2013)\citenamefont
  {Hergert}, \citenamefont {Binder}, \citenamefont {Calci}, \citenamefont
  {Langhammer},\ and\ \citenamefont {Roth}}]{hergert13_1718}%
  \BibitemOpen
  \bibfield  {author} {\bibinfo {author} {\bibfnamefont {H.}~\bibnamefont
  {Hergert}}, \bibinfo {author} {\bibfnamefont {S.}~\bibnamefont {Binder}},
  \bibinfo {author} {\bibfnamefont {A.}~\bibnamefont {Calci}}, \bibinfo
  {author} {\bibfnamefont {J.}~\bibnamefont {Langhammer}}, \ and\ \bibinfo
  {author} {\bibfnamefont {R.}~\bibnamefont {Roth}},\ }\href
  {http://dx.doi.org/10.1103/PhysRevLett.110.242501} {\bibfield  {journal}
  {\bibinfo  {journal} {Phys. Rev. Lett.}\ }\textbf {\bibinfo {volume} {110}},\
  \bibinfo {pages} {242501} (\bibinfo {year} {2013})}\BibitemShut {NoStop}%
\bibitem [{\citenamefont {Hergert}\ \emph {et~al.}(2014)\citenamefont
  {Hergert}, \citenamefont {Bogner}, \citenamefont {Morris}, \citenamefont
  {Binder}, \citenamefont {Calci}, \citenamefont {Langhammer},\ and\
  \citenamefont {Roth}}]{hergert14_1719}%
  \BibitemOpen
  \bibfield  {author} {\bibinfo {author} {\bibfnamefont {H.}~\bibnamefont
  {Hergert}}, \bibinfo {author} {\bibfnamefont {S.~K.}\ \bibnamefont {Bogner}},
  \bibinfo {author} {\bibfnamefont {T.~D.}\ \bibnamefont {Morris}}, \bibinfo
  {author} {\bibfnamefont {S.}~\bibnamefont {Binder}}, \bibinfo {author}
  {\bibfnamefont {A.}~\bibnamefont {Calci}}, \bibinfo {author} {\bibfnamefont
  {J.}~\bibnamefont {Langhammer}}, \ and\ \bibinfo {author} {\bibfnamefont
  {R.}~\bibnamefont {Roth}},\ }\href
  {http://dx.doi.org/10.1103/PhysRevC.90.041302} {\bibfield  {journal}
  {\bibinfo  {journal} {Phys. Rev. C}\ }\textbf {\bibinfo {volume} {90}},\
  \bibinfo {pages} {041302(R)} (\bibinfo {year} {2014})}\BibitemShut {NoStop}%
\bibitem [{\citenamefont {Bogner}\ \emph {et~al.}(2014)\citenamefont {Bogner},
  \citenamefont {Hergert}, \citenamefont {Holt}, \citenamefont {Schwenk},
  \citenamefont {Binder}, \citenamefont {Calci}, \citenamefont {Langhammer},\
  and\ \citenamefont {Roth}}]{bogner14_1156}%
  \BibitemOpen
  \bibfield  {author} {\bibinfo {author} {\bibfnamefont {S.~K.}\ \bibnamefont
  {Bogner}}, \bibinfo {author} {\bibfnamefont {H.}~\bibnamefont {Hergert}},
  \bibinfo {author} {\bibfnamefont {J.~D.}\ \bibnamefont {Holt}}, \bibinfo
  {author} {\bibfnamefont {A.}~\bibnamefont {Schwenk}}, \bibinfo {author}
  {\bibfnamefont {S.}~\bibnamefont {Binder}}, \bibinfo {author} {\bibfnamefont
  {A.}~\bibnamefont {Calci}}, \bibinfo {author} {\bibfnamefont
  {J.}~\bibnamefont {Langhammer}}, \ and\ \bibinfo {author} {\bibfnamefont
  {R.}~\bibnamefont {Roth}},\ }\href
  {https://dx.doi.org/10.1103/PhysRevLett.113.142501} {\bibfield  {journal}
  {\bibinfo  {journal} {Phys. Rev. Lett.}\ }\textbf {\bibinfo {volume} {113}},\
  \bibinfo {pages} {142501} (\bibinfo {year} {2014})}\BibitemShut {NoStop}%
\bibitem [{\citenamefont {Cipollone}\ \emph {et~al.}(2013)\citenamefont
  {Cipollone}, \citenamefont {Barbieri},\ and\ \citenamefont
  {Navr\'atil}}]{cipollone13_1857}%
  \BibitemOpen
  \bibfield  {author} {\bibinfo {author} {\bibfnamefont {A.}~\bibnamefont
  {Cipollone}}, \bibinfo {author} {\bibfnamefont {C.}~\bibnamefont {Barbieri}},
  \ and\ \bibinfo {author} {\bibfnamefont {P.}~\bibnamefont {Navr\'atil}},\
  }\href {http://dx.doi.org/10.1103/PhysRevLett.111.062501} {\bibfield
  {journal} {\bibinfo  {journal} {Phys. Rev. Lett.}\ }\textbf {\bibinfo
  {volume} {111}},\ \bibinfo {pages} {062501} (\bibinfo {year}
  {2013})}\BibitemShut {NoStop}%
\bibitem [{\citenamefont {Jansen}\ \emph {et~al.}(2014)\citenamefont {Jansen},
  \citenamefont {Engel}, \citenamefont {Hagen}, \citenamefont {Navr\'atil},\
  and\ \citenamefont {Signoracci}}]{jansen14_1150}%
  \BibitemOpen
  \bibfield  {author} {\bibinfo {author} {\bibfnamefont {G.~R.}\ \bibnamefont
  {Jansen}}, \bibinfo {author} {\bibfnamefont {J.}~\bibnamefont {Engel}},
  \bibinfo {author} {\bibfnamefont {G.}~\bibnamefont {Hagen}}, \bibinfo
  {author} {\bibfnamefont {P.}~\bibnamefont {Navr\'atil}}, \ and\ \bibinfo
  {author} {\bibfnamefont {A.}~\bibnamefont {Signoracci}},\ }\href
  {https://dx.doi.org/10.1103/PhysRevLett.113.142502} {\bibfield  {journal}
  {\bibinfo  {journal} {Phys. Rev. Lett.}\ }\textbf {\bibinfo {volume} {113}},\
  \bibinfo {pages} {142502} (\bibinfo {year} {2014})}\BibitemShut {NoStop}%
\bibitem [{\citenamefont {Epelbaum}\ \emph {et~al.}(2014)\citenamefont
  {Epelbaum}, \citenamefont {Krebs}, \citenamefont {L\"ahde}, \citenamefont
  {Lee}, \citenamefont {Mei{\ss}ner},\ and\ \citenamefont
  {Rupak}}]{epelbaum14_1893}%
  \BibitemOpen
  \bibfield  {author} {\bibinfo {author} {\bibfnamefont {E.}~\bibnamefont
  {Epelbaum}}, \bibinfo {author} {\bibfnamefont {H.}~\bibnamefont {Krebs}},
  \bibinfo {author} {\bibfnamefont {T.~A.}\ \bibnamefont {L\"ahde}}, \bibinfo
  {author} {\bibfnamefont {D.}~\bibnamefont {Lee}}, \bibinfo {author}
  {\bibfnamefont {U.}~\bibnamefont {Mei{\ss}ner}}, \ and\ \bibinfo {author}
  {\bibfnamefont {G.}~\bibnamefont {Rupak}},\ }\href
  {https://doi.org/10.1103/PhysRevLett.112.102501} {\bibfield  {journal}
  {\bibinfo  {journal} {Phys. Rev. Lett.}\ }\textbf {\bibinfo {volume} {112}},\
  \bibinfo {pages} {102501} (\bibinfo {year} {2014})}\BibitemShut {NoStop}%
\bibitem [{\citenamefont {Binder}\ \emph {et~al.}(2014)\citenamefont {Binder},
  \citenamefont {Langhammer}, \citenamefont {Calci},\ and\ \citenamefont
  {Roth}}]{binder14_1882}%
  \BibitemOpen
  \bibfield  {author} {\bibinfo {author} {\bibfnamefont {S.}~\bibnamefont
  {Binder}}, \bibinfo {author} {\bibfnamefont {J.}~\bibnamefont {Langhammer}},
  \bibinfo {author} {\bibfnamefont {A.}~\bibnamefont {Calci}}, \ and\ \bibinfo
  {author} {\bibfnamefont {R.}~\bibnamefont {Roth}},\ }\href
  {http://dx.doi.org/10.1016/j.physletb.2014.07.010} {\bibfield  {journal}
  {\bibinfo  {journal} {Phys. Lett. B}\ }\textbf {\bibinfo {volume} {736}},\
  \bibinfo {pages} {119} (\bibinfo {year} {2014})}\BibitemShut {NoStop}%
\bibitem [{\citenamefont {Cipollone}\ \emph {et~al.}(2015)\citenamefont
  {Cipollone}, \citenamefont {Barbieri},\ and\ \citenamefont
  {Navr\'atil}}]{cipollone15_1894}%
  \BibitemOpen
  \bibfield  {author} {\bibinfo {author} {\bibfnamefont {A.}~\bibnamefont
  {Cipollone}}, \bibinfo {author} {\bibfnamefont {C.}~\bibnamefont {Barbieri}},
  \ and\ \bibinfo {author} {\bibfnamefont {P.}~\bibnamefont {Navr\'atil}},\
  }\href {https://doi.org/10.1103/PhysRevC.92.014306} {\bibfield  {journal}
  {\bibinfo  {journal} {Phys. Rev. C}\ }\textbf {\bibinfo {volume} {92}},\
  \bibinfo {pages} {014306} (\bibinfo {year} {2015})}\BibitemShut {NoStop}%
\bibitem [{\citenamefont {Hergert}(2017)}]{hergert17_1873}%
  \BibitemOpen
  \bibfield  {author} {\bibinfo {author} {\bibfnamefont {H.}~\bibnamefont
  {Hergert}},\ }\href {http://dx.doi.org/10.1088/1402-4896/92/2/023002}
  {\bibfield  {journal} {\bibinfo  {journal} {Phys. Scr.}\ }\textbf {\bibinfo
  {volume} {92}},\ \bibinfo {pages} {023002} (\bibinfo {year}
  {2017})}\BibitemShut {NoStop}%
\bibitem [{\citenamefont {Stroberg}\ \emph {et~al.}(2017)\citenamefont
  {Stroberg}, \citenamefont {Calci}, \citenamefont {Hergert}, \citenamefont
  {Holt}, \citenamefont {Bogner}, \citenamefont {Roth},\ and\ \citenamefont
  {Schwenk}}]{stroberg17_1807}%
  \BibitemOpen
  \bibfield  {author} {\bibinfo {author} {\bibfnamefont {S.~R.}\ \bibnamefont
  {Stroberg}}, \bibinfo {author} {\bibfnamefont {A.}~\bibnamefont {Calci}},
  \bibinfo {author} {\bibfnamefont {H.}~\bibnamefont {Hergert}}, \bibinfo
  {author} {\bibfnamefont {J.~D.}\ \bibnamefont {Holt}}, \bibinfo {author}
  {\bibfnamefont {S.~K.}\ \bibnamefont {Bogner}}, \bibinfo {author}
  {\bibfnamefont {R.}~\bibnamefont {Roth}}, \ and\ \bibinfo {author}
  {\bibfnamefont {A.}~\bibnamefont {Schwenk}},\ }\href
  {http://dx.doi.org/10.1103/PhysRevLett.118.032502} {\bibfield  {journal}
  {\bibinfo  {journal} {Phys. Rev. Lett.}\ }\textbf {\bibinfo {volume} {118}},\
  \bibinfo {pages} {032502} (\bibinfo {year} {2017})}\BibitemShut {NoStop}%
\bibitem [{\citenamefont {Tichai}\ \emph {et~al.}(2017)\citenamefont {Tichai},
  \citenamefont {Gebrerufael},\ and\ \citenamefont {Roth}}]{arxivTichai17}%
  \BibitemOpen
  \bibfield  {author} {\bibinfo {author} {\bibfnamefont {A.}~\bibnamefont
  {Tichai}}, \bibinfo {author} {\bibfnamefont {E.}~\bibnamefont {Gebrerufael}},
  \ and\ \bibinfo {author} {\bibfnamefont {R.}~\bibnamefont {Roth}},\
  }\href@noop {} {\enquote {\bibinfo {title} {Open-shell nuclei from no-core
  shell model with perturbative improvement},}\ }\bibinfo {howpublished}
  {\url{https://arxiv.org/abs/1703.05664}} (\bibinfo {year} {2017})\BibitemShut
  {NoStop}%
\bibitem [{\citenamefont {Hebeler}\ \emph {et~al.}(2015)\citenamefont
  {Hebeler}, \citenamefont {Holt}, \citenamefont {Men\'endez},\ and\
  \citenamefont {Schwenk}}]{hebeler15_1872}%
  \BibitemOpen
  \bibfield  {author} {\bibinfo {author} {\bibfnamefont {K.}~\bibnamefont
  {Hebeler}}, \bibinfo {author} {\bibfnamefont {J.~D.}\ \bibnamefont {Holt}},
  \bibinfo {author} {\bibfnamefont {J.}~\bibnamefont {Men\'endez}}, \ and\
  \bibinfo {author} {\bibfnamefont {A.}~\bibnamefont {Schwenk}},\ }\href
  {https://doi.org/10.1146/annurev-nucl-102313-025446} {\bibfield  {journal}
  {\bibinfo  {journal} {Annu. Rev. Nucl. Part. Sci.}\ }\textbf {\bibinfo
  {volume} {65}},\ \bibinfo {pages} {457} (\bibinfo {year} {2015})}\BibitemShut
  {NoStop}%
\bibitem [{\citenamefont {Lapoux}\ \emph {et~al.}(2016)\citenamefont {Lapoux},
  \citenamefont {Som\`a}, \citenamefont {Barbieri}, \citenamefont {Hergert},
  \citenamefont {Holt},\ and\ \citenamefont {Stroberg}}]{lapoux16_1782}%
  \BibitemOpen
  \bibfield  {author} {\bibinfo {author} {\bibfnamefont {V.}~\bibnamefont
  {Lapoux}}, \bibinfo {author} {\bibfnamefont {V.}~\bibnamefont {Som\`a}},
  \bibinfo {author} {\bibfnamefont {C.}~\bibnamefont {Barbieri}}, \bibinfo
  {author} {\bibfnamefont {H.}~\bibnamefont {Hergert}}, \bibinfo {author}
  {\bibfnamefont {J.~D.}\ \bibnamefont {Holt}}, \ and\ \bibinfo {author}
  {\bibfnamefont {S.~R.}\ \bibnamefont {Stroberg}},\ }\href
  {http://dx.doi.org/10.1103/PhysRevLett.117.052501} {\bibfield  {journal}
  {\bibinfo  {journal} {Phys. Rev. Lett.}\ }\textbf {\bibinfo {volume} {117}},\
  \bibinfo {pages} {052501} (\bibinfo {year} {2016})}\BibitemShut {NoStop}%
\bibitem [{\citenamefont {{H. Hergert, private
  communication}}(2016)}]{privcom}%
  \BibitemOpen
  \bibfield  {author} {\bibinfo {author} {\bibnamefont {{H. Hergert, private
  communication}}},\ }\href@noop {} {} (\bibinfo {year} {2016})\BibitemShut
  {NoStop}%
\bibitem [{\citenamefont {Michel}\ \emph {et~al.}(2010)\citenamefont {Michel},
  \citenamefont {Nazarewicz}, \citenamefont {Oko{\l}owicz},\ and\ \citenamefont
  {P{\l}oszajczak}}]{michel10_4}%
  \BibitemOpen
  \bibfield  {author} {\bibinfo {author} {\bibfnamefont {N.}~\bibnamefont
  {Michel}}, \bibinfo {author} {\bibfnamefont {W.}~\bibnamefont {Nazarewicz}},
  \bibinfo {author} {\bibfnamefont {J.}~\bibnamefont {Oko{\l}owicz}}, \ and\
  \bibinfo {author} {\bibfnamefont {M.}~\bibnamefont {P{\l}oszajczak}},\ }\href
  {https://dx.doi.org/10.1088/0954-3899/37/6/064042} {\bibfield  {journal}
  {\bibinfo  {journal} {J. Phys. G}\ }\textbf {\bibinfo {volume} {37}},\
  \bibinfo {pages} {064042} (\bibinfo {year} {2010})}\BibitemShut {NoStop}%
\bibitem [{\citenamefont {Michel}\ \emph {et~al.}(2009)\citenamefont {Michel},
  \citenamefont {Nazarewicz}, \citenamefont {P{\l}oszajczak},\ and\
  \citenamefont {Vertse}}]{michel09_2}%
  \BibitemOpen
  \bibfield  {author} {\bibinfo {author} {\bibfnamefont {N.}~\bibnamefont
  {Michel}}, \bibinfo {author} {\bibfnamefont {W.}~\bibnamefont {Nazarewicz}},
  \bibinfo {author} {\bibfnamefont {M.}~\bibnamefont {P{\l}oszajczak}}, \ and\
  \bibinfo {author} {\bibfnamefont {T.}~\bibnamefont {Vertse}},\ }\href
  {https://dx.doi.org/10.1088/0954-3899/36/1/013101} {\bibfield  {journal}
  {\bibinfo  {journal} {J. Phys. G}\ }\textbf {\bibinfo {volume} {36}},\
  \bibinfo {pages} {013101} (\bibinfo {year} {2009})}\BibitemShut {NoStop}%
\bibitem [{\citenamefont {Berggren}(1968)}]{berggren68_32}%
  \BibitemOpen
  \bibfield  {author} {\bibinfo {author} {\bibfnamefont {T.}~\bibnamefont
  {Berggren}},\ }\href {https://dx.doi.org/10.1016/0375-9474(68)90593-9}
  {\bibfield  {journal} {\bibinfo  {journal} {Nucl. Phys. A}\ }\textbf
  {\bibinfo {volume} {109}},\ \bibinfo {pages} {265} (\bibinfo {year}
  {1968})}\BibitemShut {NoStop}%
\bibitem [{\citenamefont {Berggren}(1982)}]{berggren82_28}%
  \BibitemOpen
  \bibfield  {author} {\bibinfo {author} {\bibfnamefont {T.}~\bibnamefont
  {Berggren}},\ }\href {https://dx.doi.org/10.1016/0375-9474(82)90519-X}
  {\bibfield  {journal} {\bibinfo  {journal} {Nucl. Phys. A}\ }\textbf
  {\bibinfo {volume} {389}},\ \bibinfo {pages} {261} (\bibinfo {year}
  {1982})}\BibitemShut {NoStop}%
\bibitem [{\citenamefont {Rotureau}\ \emph {et~al.}(2006)\citenamefont
  {Rotureau}, \citenamefont {Michel}, \citenamefont {Nazarewicz}, \citenamefont
  {P{\l}oszajczak},\ and\ \citenamefont {Dukelsky}}]{rotureau06_15}%
  \BibitemOpen
  \bibfield  {author} {\bibinfo {author} {\bibfnamefont {J.}~\bibnamefont
  {Rotureau}}, \bibinfo {author} {\bibfnamefont {N.}~\bibnamefont {Michel}},
  \bibinfo {author} {\bibfnamefont {W.}~\bibnamefont {Nazarewicz}}, \bibinfo
  {author} {\bibfnamefont {M.}~\bibnamefont {P{\l}oszajczak}}, \ and\ \bibinfo
  {author} {\bibfnamefont {J.}~\bibnamefont {Dukelsky}},\ }\href
  {https://dx.doi.org/10.1103/PhysRevLett.97.110603} {\bibfield  {journal}
  {\bibinfo  {journal} {Phys. Rev. Lett.}\ }\textbf {\bibinfo {volume} {97}},\
  \bibinfo {pages} {110603} (\bibinfo {year} {2006})}\BibitemShut {NoStop}%
\bibitem [{\citenamefont {Rotureau}\ \emph {et~al.}(2009)\citenamefont
  {Rotureau}, \citenamefont {Michel}, \citenamefont {Nazarewicz}, \citenamefont
  {P{\l}oszajczak},\ and\ \citenamefont {Dukelsky}}]{rotureau09_140}%
  \BibitemOpen
  \bibfield  {author} {\bibinfo {author} {\bibfnamefont {J.}~\bibnamefont
  {Rotureau}}, \bibinfo {author} {\bibfnamefont {N.}~\bibnamefont {Michel}},
  \bibinfo {author} {\bibfnamefont {W.}~\bibnamefont {Nazarewicz}}, \bibinfo
  {author} {\bibfnamefont {M.}~\bibnamefont {P{\l}oszajczak}}, \ and\ \bibinfo
  {author} {\bibfnamefont {J.}~\bibnamefont {Dukelsky}},\ }\href
  {https://dx.doi.org/10.1103/PhysRevC.79.014304} {\bibfield  {journal}
  {\bibinfo  {journal} {Phys. Rev. C}\ }\textbf {\bibinfo {volume} {79}},\
  \bibinfo {pages} {014304} (\bibinfo {year} {2009})}\BibitemShut {NoStop}%
\bibitem [{ens()}]{ensdf}%
  \BibitemOpen
  \href@noop {} {}\bibinfo {howpublished}
  {\url{http://www.nndc.bnl.gov/ensdf}}\BibitemShut {NoStop}%
\bibitem [{\citenamefont {Grigorenko}\ \emph {et~al.}(2013)\citenamefont
  {Grigorenko}, \citenamefont {Mukha},\ and\ \citenamefont
  {Zhukov}}]{grigorenko13_1219}%
  \BibitemOpen
  \bibfield  {author} {\bibinfo {author} {\bibfnamefont {L.~V.}\ \bibnamefont
  {Grigorenko}}, \bibinfo {author} {\bibfnamefont {I.~G.}\ \bibnamefont
  {Mukha}}, \ and\ \bibinfo {author} {\bibfnamefont {M.~V.}\ \bibnamefont
  {Zhukov}},\ }\href {https://dx.doi.org/10.1103/PhysRevLett.111.042501}
  {\bibfield  {journal} {\bibinfo  {journal} {Phys. Rev. Lett.}\ }\textbf
  {\bibinfo {volume} {111}},\ \bibinfo {pages} {042501} (\bibinfo {year}
  {2013})}\BibitemShut {NoStop}%
\bibitem [{\citenamefont {Grigorenko}\ and\ \citenamefont
  {Zhukov}(2015)}]{grigorenko15_1874}%
  \BibitemOpen
  \bibfield  {author} {\bibinfo {author} {\bibfnamefont {L.~V.}\ \bibnamefont
  {Grigorenko}}\ and\ \bibinfo {author} {\bibfnamefont {M.~V.}\ \bibnamefont
  {Zhukov}},\ }\href {http://dx.doi.org/10.1103/PhysRevC.91.064617} {\bibfield
  {journal} {\bibinfo  {journal} {Phys. Rev. C}\ }\textbf {\bibinfo {volume}
  {91}},\ \bibinfo {pages} {064617} (\bibinfo {year} {2015})}\BibitemShut
  {NoStop}%
\bibitem [{\citenamefont {Hove}\ \emph {et~al.}(2017)\citenamefont {Hove},
  \citenamefont {Garrido}, \citenamefont {Sarriguren}, \citenamefont {Fedorov},
  \citenamefont {Fynbo}, \citenamefont {Jensen},\ and\ \citenamefont
  {Zinner}}]{hove17_1893}%
  \BibitemOpen
  \bibfield  {author} {\bibinfo {author} {\bibfnamefont {D.}~\bibnamefont
  {Hove}}, \bibinfo {author} {\bibfnamefont {E.}~\bibnamefont {Garrido}},
  \bibinfo {author} {\bibfnamefont {P.}~\bibnamefont {Sarriguren}}, \bibinfo
  {author} {\bibfnamefont {D.~V.}\ \bibnamefont {Fedorov}}, \bibinfo {author}
  {\bibfnamefont {H.~O.~U.}\ \bibnamefont {Fynbo}}, \bibinfo {author}
  {\bibfnamefont {A.~S.}\ \bibnamefont {Jensen}}, \ and\ \bibinfo {author}
  {\bibfnamefont {N.~T.}\ \bibnamefont {Zinner}},\ }\href
  {https://doi.org/10.1103/PhysRevC.95.061301} {\bibfield  {journal} {\bibinfo
  {journal} {Phys. Rev. C}\ }\textbf {\bibinfo {volume} {95}},\ \bibinfo
  {pages} {061301(R)} (\bibinfo {year} {2017})}\BibitemShut {NoStop}%
\bibitem [{\citenamefont {Furutani}\ \emph {et~al.}(1978)\citenamefont
  {Furutani}, \citenamefont {Horiuchi},\ and\ \citenamefont
  {Tamagaki}}]{furutani78_1012}%
  \BibitemOpen
  \bibfield  {author} {\bibinfo {author} {\bibfnamefont {H.}~\bibnamefont
  {Furutani}}, \bibinfo {author} {\bibfnamefont {H.}~\bibnamefont {Horiuchi}},
  \ and\ \bibinfo {author} {\bibfnamefont {R.}~\bibnamefont {Tamagaki}},\
  }\href {https://dx.doi.org/10.1143/PTP.60.307} {\bibfield  {journal}
  {\bibinfo  {journal} {Prog. Theor. Phys.}\ }\textbf {\bibinfo {volume}
  {60}},\ \bibinfo {pages} {307} (\bibinfo {year} {1978})}\BibitemShut
  {NoStop}%
\bibitem [{\citenamefont {Furutani}\ \emph {et~al.}(1979)\citenamefont
  {Furutani}, \citenamefont {Horiuchi},\ and\ \citenamefont
  {Tamagaki}}]{furutani79_1013}%
  \BibitemOpen
  \bibfield  {author} {\bibinfo {author} {\bibfnamefont {H.}~\bibnamefont
  {Furutani}}, \bibinfo {author} {\bibfnamefont {H.}~\bibnamefont {Horiuchi}},
  \ and\ \bibinfo {author} {\bibfnamefont {R.}~\bibnamefont {Tamagaki}},\
  }\href {https://dx.doi.org/10.1143/PTP.62.981} {\bibfield  {journal}
  {\bibinfo  {journal} {Prog. Theor. Phys.}\ }\textbf {\bibinfo {volume}
  {62}},\ \bibinfo {pages} {981} (\bibinfo {year} {1979})}\BibitemShut
  {NoStop}%
\bibitem [{\citenamefont {Catara}\ \emph {et~al.}(1984)\citenamefont {Catara},
  \citenamefont {Insolia}, \citenamefont {Maglione},\ and\ \citenamefont
  {Vitturi}}]{catara84_1880}%
  \BibitemOpen
  \bibfield  {author} {\bibinfo {author} {\bibfnamefont {F.}~\bibnamefont
  {Catara}}, \bibinfo {author} {\bibfnamefont {A.}~\bibnamefont {Insolia}},
  \bibinfo {author} {\bibfnamefont {E.}~\bibnamefont {Maglione}}, \ and\
  \bibinfo {author} {\bibfnamefont {A.}~\bibnamefont {Vitturi}},\ }\href
  {https://doi.org/10.1103/PhysRevC.29.1091} {\bibfield  {journal} {\bibinfo
  {journal} {Phys. Rev. C}\ }\textbf {\bibinfo {volume} {29}},\ \bibinfo
  {pages} {1091} (\bibinfo {year} {1984})}\BibitemShut {NoStop}%
\bibitem [{\citenamefont {Pillet}\ \emph {et~al.}(2007)\citenamefont {Pillet},
  \citenamefont {Sandulescu},\ and\ \citenamefont {Schuck}}]{pillet07_1879}%
  \BibitemOpen
  \bibfield  {author} {\bibinfo {author} {\bibfnamefont {N.}~\bibnamefont
  {Pillet}}, \bibinfo {author} {\bibfnamefont {N.}~\bibnamefont {Sandulescu}},
  \ and\ \bibinfo {author} {\bibfnamefont {P.}~\bibnamefont {Schuck}},\ }\href
  {http://dx.doi.org/10.1103/PhysRevC.76.024310} {\bibfield  {journal}
  {\bibinfo  {journal} {Phys. Rev. C}\ }\textbf {\bibinfo {volume} {76}},\
  \bibinfo {pages} {024310} (\bibinfo {year} {2007})}\BibitemShut {NoStop}%
\bibitem [{\citenamefont {Hagino}\ and\ \citenamefont
  {Sagawa}(2014)}]{hagino14_1160}%
  \BibitemOpen
  \bibfield  {author} {\bibinfo {author} {\bibfnamefont {K.}~\bibnamefont
  {Hagino}}\ and\ \bibinfo {author} {\bibfnamefont {H.}~\bibnamefont
  {Sagawa}},\ }\href {https://dx.doi.org/10.1103/PhysRevC.89.014331} {\bibfield
   {journal} {\bibinfo  {journal} {Phys. Rev. C}\ }\textbf {\bibinfo {volume}
  {89}},\ \bibinfo {pages} {014331} (\bibinfo {year} {2014})}\BibitemShut
  {NoStop}%
\bibitem [{\citenamefont {Hagino}\ and\ \citenamefont
  {Sagawa}(2016)}]{hagino16_1881}%
  \BibitemOpen
  \bibfield  {author} {\bibinfo {author} {\bibfnamefont {K.}~\bibnamefont
  {Hagino}}\ and\ \bibinfo {author} {\bibfnamefont {S.}~\bibnamefont
  {Sagawa}},\ }\href {http://dx.doi.org/10.1007/s00601-015-1027-3} {\bibfield
  {journal} {\bibinfo  {journal} {Few-Body Syst.}\ }\textbf {\bibinfo {volume}
  {57}},\ \bibinfo {pages} {185} (\bibinfo {year} {2016})}\BibitemShut
  {NoStop}%
\end{thebibliography}

%

\end{document}